\documentclass[11pt]{article}
\usepackage[top=1in, bottom=1in, left=1in, right=1in]{geometry}
\usepackage[labelfont=bf]{caption}
\usepackage{graphicx}
\usepackage{enumitem}

\begin{document}

\noindent{\Huge\textbf{\textsf{The Sun's magnetic midlife crisis}}}
\vspace*{11pt}

\noindent{\Large\textbf{\textsf{Travis S. Metcalfe}}}
\vspace*{11pt}

\noindent{\textsf{Old Sun-like stars observed by the Kepler Space Telescope spin faster than astronomers expected. Apparently, they experience a dramatic shift in their rotation and magnetism at about middle age.}}
\vspace*{11pt}

\noindent The Sun is just one of a hundred billion stars in the Milky Way galaxy. Our front-row 
seat on Earth allows us to observe it in much greater detail than we can for other stars. 
However, those observations provide only one snapshot in the life story of stars like the 
Sun. To piece together the entire tale, astronomers need to study other stars that are 
younger and older. Their observations have revealed something unexpected that never could 
have come to light from studies of our Sun in isolation.

\vspace*{-11pt}\section*{\textsf{Magnetic braking}}\vspace*{-6pt}

Almost half a century ago, astrophysicist Andrew Skumanich noticed that rotation and 
magnetism in stars are intricately linked. Based on observations of the Sun and a few 
young star clusters for which stellar ages could be estimated, he proposed that both the 
rotation rate and the magnetic field strength in stars are inversely proportional to the 
square root of age.

During a total solar eclipse (panel a of the figure), you can see a clue as to why 
rotation and magnetism are related: the solar wind, a steady stream of material that 
emanates from the Sun's surface. The charged particles in the wind follow the magnetic 
field lines outward, and because the field lines rotate with the Sun, the outgoing 
particles gain angular momentum at the Sun's expense. On reaching a sufficient distance, 
the particles break free and carry away their angular momentum. The resulting gradual 
slowing of the Sun's rotation is called magnetic braking. Rotation, in turn, modifies the 
magnetic field because the Sun rotates faster at the equator than near the poles, a 
phenomenon known as differential rotation. The resulting shear wraps up the large-scale 
field and creates small loops that eventually appear as sunspots. Over time, rotation and 
magnetism diminish together, each feeding off the other.

By the late 1980s, astronomers had recognized that magnetic braking is stronger in more 
rapidly rotating stars. Thus, even though stars are formed with a range of initial 
rotation rates, the rates tend to converge to a single, mass-dependent value. For stars 
that, like the Sun, have outer convection zones where temperature gradients drive the 
rise and fall of fluid elements, the convergence time is roughly 500 million years. 
However, the evidence for the above scenario relies on studies of rotation in members of 
young star clusters; until recently the only older star available for testing the ideas 
was the Sun.

\vspace*{-11pt}\section*{\textsf{Old but spry}}\vspace*{-6pt}

The Kepler Space Telescope changed our understanding of how rotation and magnetism evolve 
in Sun-like stars. After its launch in 2009, Kepler spent four years monitoring thousands 
of stars and a few clusters. Rotation of the stars in clusters with ages up to 2.5 
billion years agreed with previous expectations. But a different behavior was revealed by 
some of the isolated stars, whose ages were determined from asteroseismology, essentially 
analyses of brightness oscillations. The younger stars agreed with the clusters, but the 
older stars rotated more quickly than expected. Beyond middle age, the angular momentum 
of stars no longer appeared to be decreasing over time.

The anomalous rotation became significant near 4--5 billion years for so-called G-type 
stars, the class that includes our Sun, but it appeared after 2--3 billion years for 
hotter F-type stars and after 6--7 billion years for cooler K-type stars. That dependence 
on the temperature suggested a link to the convection zones, whose fluid motion is 
thought to contribute to the generation of magnetic fields. Specifically, magnetic 
braking appears to shut down at a constant Rossby number, defined as the ratio of the 
rotation period to the turnover time for fluid elements in the convection zone. Cooler 
stars have deeper convection zones with longer turnover times, so they would continue 
braking for longer before reaching the slower rotation required to yield the critical 
Rossby number. In a 2016 Nature paper, Jennifer van Saders and colleagues reproduced the 
observations with rotational evolution models based on that idea.

Ground-based spectroscopic observations of the Kepler stars revealed that the critical 
Rossby number could also be interpreted in terms of magnetic fields. It turns out that 
two specific spectral lines from calcium ions can serve as a useful proxy for the 
strength and fractional area covered by magnetic fields. The critical Rossby number 
corresponds to a specific luminosity of calcium emission from the chromosphere, the part 
of the solar atmosphere just above the light-emitting photosphere. Stars with different 
surface temperatures take different amounts of time to reach that level of chromospheric 
emission. Note, though, that as the star ages, chromospheric emission continues to 
decrease before eventually settling down to a constant value; in that regard it differs 
from stellar rotation, which is locked in place by the shutdown of magnetic braking at 
the critical Rossby value.

As the star's rotation rate decreases and the Rossby number approaches the critical 
value, the global magnetic field is concentrated into smaller spatial scales and is a 
much less efficient means for shedding angular momentum. How might that concentration 
come about? As the rotation period becomes comparable to the convective turnover time, 
the imprint of Coriolis forces on the convective patterns is diminished. The reduced 
effect might naturally lead to a change in the character of differential rotation that 
modifies the dominant scale of the global magnetic field.

\vspace*{-11pt}\section*{\textsf{Demise of the solar cycle}}\vspace*{-6pt}

The revised picture of rotational and magnetic evolution just described not only explains 
the new observations from Kepler, it also addresses a long-standing puzzle about the 
11-year sunspot cycle. The plot in panel b of the figure shows an updated version of a 
diagram originally published by Erika B\"ohm-Vitense more than a decade ago. It presents 
observations of magnetic cycles and rotation periods for stars studied by the Mount 
Wilson Observatory HK Project, which started in 1966 and ran for more than 35 years. 
B\"ohm-Vitense noted two distinct relationships between the rotation period and the 
length of the stellar cycle. Most significantly, she found that the 11-year solar cycle 
appeared to fall between the two stellar sequences.

The updated version of the diagram colors the points to indicate the surface temperature 
of each star. At the critical Rossby number corresponding to the shutdown of magnetic 
braking, stars deviate from one of the relationships observed by B\"ohm-Vitense; 
eventually the stars will evolve to have constant chromospheric emission. Stars that 
previously seemed to be outliers, including the Sun, 5 Serpentis (5 Ser), Alpha Centauri 
A ($\alpha$ Cen A), and 94 Aquarii A (94 Aqr A), can now be understood as transitioning 
from predominantly large-scale to smaller-scale global magnetic fields and longer 
magnetic cycles. The other stars indicated on the diagram are 18 Scorpii (18 Sco), 15 
Coronae Borealis (15 CrB), 16 Cygni B (16 Cyg B), and 31 Aquilae (31 Aql).

Future tests of those new ideas about magnetic evolution will come from ground-based 
chromospheric emission measurements of Kepler targets that span the putative magnetic 
transition and from asteroseismology with the newly launched Transiting Exoplanet Survey 
Satellite to determine precise masses and ages for the bright stars with known magnetic 
cycles. Additional insights might also come from more difficult measurements of 
differential rotation and from reconstructions of the magnetic field geometry obtained 
from analysis of the polarization of stellar light.

 \begin{figure*}[!t]
 \centering\includegraphics[angle=270,width=5.25in]{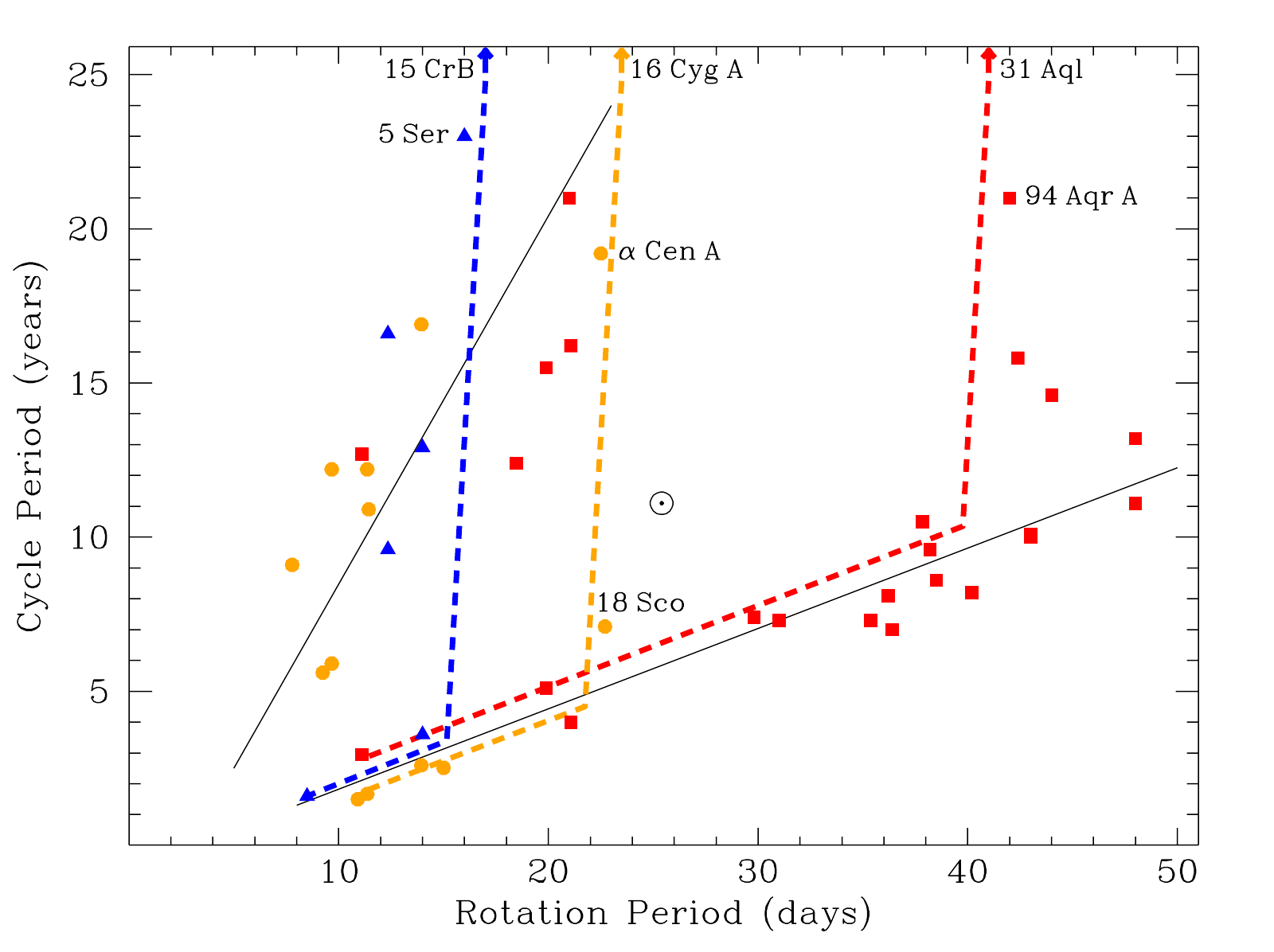}
 \caption{\sf THE DYNAMIC SUN. A solar eclipse (a) makes visible the solar wind, material 
that follows magnetic field lines as it leaves the Sun's surface. (Photo courtesy of 
Miroslav Druckm\"uller.) (b)~In this plot relating the length of the stellar magnetic 
cycle to the rotation period, the solid black lines show the two distinct relationships 
observed by Erika B\"ohm-Vitense about a decade ago. Points are colored according to the 
star's surface temperature; hotter F-type stars are blue, G-type stars like the Sun are 
orange, and cooler K-type stars are red. The colored lines show representative evolution 
paths that deviate from B\"ohm-Vitense's lower relationship when the rotation period 
corresponds to the critical Rossby number heralding the shutdown of magnetic braking. 
(For a possible interpretation of the upper relationship, see T. Metcalfe, J. van Saders, 
https://arxiv.org/abs/1705.09668.) Eventually stars evolve to have constant chromospheric 
emission, which appears to indicate that they have completed the transition from 
predominantly large-scale to smaller-scale global magnetic fields. The Sun, indicated by 
the $\odot$ symbol, falls to the right of the orange line because it is slightly cooler 
than the other indicated stars and thus reaches the critical Rossby number at a longer 
rotation period. As older stars lie higher along the colored lines than younger stars, 
the observations suggest that the 11-year sunspot cycle may grow longer before it 
eventually disappears. The three stars at the top of the graph are indicated with arrows 
because they show no detectable cycle period during 25 years of observations.}
 \end{figure*}

\vspace*{-11pt}\section*{\textsf{Additional Resources}}\vspace*{-6pt}
\begin{itemize}[leftmargin=1em]
\item A. Skumanich, ``Time scales for Ca II emission decay, rotational braking, and lithium depletion,'' Astrophys. J. 171, 565 (1972).
\item J. L. van Saders et al., ``Weakened magnetic braking as the origin of anomalously rapid rotation in old ﬁeld stars,'' Nature 529, 181 (2016).
\item E. Böhm-Vitense, ``Chromospheric activity in G and K main-sequence stars, and what it tells us about stellar dynamos,'' Astrophys. J. 657, 486 (2007).
\end{itemize}

\end{document}